\documentclass[a4paper]{jpconf}
\usepackage{graphicx}

\usepackage{citesort}

\begin{document}
\title{Auto-tuning capabilities of the ACTS track reconstruction suite}

\author{Corentin Allaire$^{1}$, Rocky Bala Garg$^{2}$, Hadrien Benjamin Grasland$^{1}$, Elyssa Frances Hofgard$^{2}$, David Rousseau$^{1}$, Rama Salahat$^{1,4}$, Andreas Salzburger$^{3}$, Lauren Alexandra Tompkins$^{2}$}

\address{$^{1}$Universit\'e Paris-Saclay, CNRS/IN2P3, IJCLab, 91405 Orsay, France}
\address{$^{2}$Stanford University, Stanford, CA 94305, USA}
\address{$^{3}$CERN, 1211 Geneva, Switzerland}
\address{$^{4}$An-Najah National University, P400 Nablus, Palestine.}

\ead{corentin.allaire@cern.ch}

\begin{abstract}
The reconstruction of charged particle trajectories is a crucial challenge of particle physics experiments as it directly impacts particle reconstruction and physics performances. To reconstruct these trajectories, different reconstruction algorithms are used sequentially. Each of these algorithms uses many configuration parameters that must be fine-tuned to properly account for the detector/experimental setup, the available CPU budget and the desired physics performance. Examples of such parameters are cut values limiting the algorithm's search space, approximations accounting for complex phenomenons, or parameters controlling algorithm performance. Until now, these parameters had to be optimised by human experts, which is inefficient and raises issues for the long-term maintainability of such algorithms. Previous experience using machine learning for particle reconstruction (such as the TrackML challenge) has shown that they can be easily adapted to different experiments by learning directly from the data. We propose to bring the same approach to the classic track reconstruction algorithms by connecting them to an agent-driven optimiser, allowing us to find the best input parameters using an iterative tuning approach. We have so far demonstrated this method on different track reconstruction algorithms within A Common Tracking Software (ACTS) framework using the Open Data Detector (ODD). These algorithms include the trajectory seed reconstruction and selection, the particle vertex reconstruction and the generation of simplified material maps used for trajectory reconstruction.
\end{abstract}

\section{Introduction}

	Charged particles trajectories reconstruction is essential to most high-energy physics experiments. A precise measurement of those trajectories allows us to properly reconstruct the primary interaction vertex, identify the particles and evaluate their momentum. Trajectory reconstruction is also one of the most computationally intensive components of event reconstruction, as it scales quadratically with the number of particles in the detector. In this context, great efforts are being deployed to optimise such algorithms' performances in terms of physics object reconstruction and computing efficiency. One option for such optimisation is tuning the different parameters of the algorithms. Those parameters range from preselection of the particles' properties to simplification of the detector effects or limits to the search space. Due to the large number of parameters for each algorithm, searching for the optimal combination can be difficult and time-consuming. To make matters worse, the optimum would change depending on the detector used, the physics studied, and the experimental conditions, demanding updates almost annually.

	In this paper, we propose using auto-tuning techniques to optimise the parameters of different tracking algorithms: constructing simplified material models for the detectors (material mapping), seeding of the tracks and reconstructing the primary vertices. Using data-driven techniques will allow us to learn the optimal set of parameters as a function of the experimental condition with minimum human input. The subsequent time savings should then allow more granular parameter optimisation that could be region-specific, signature specific... 
	
\section{ACTS and dataset}

	A Common Tracking Software (ACTS)\cite{Ai:2021ghi} is a tracking framework being developed since 2016 as an international collaboration with the goal of providing a generic, experiment-independent open-source software framework for charged particle tracks reconstruction. Since this framework is meant to be used by many different physics experiments, it gives a great opportunity to test auto-tuning with tracking algorithms on multiple configurations. The optimisation algorithms presented in this paper have been implemented in the ACTS library and are thus usable by its many users for their optimisation needs.

	The ACTS framework also implements two virtual detectors for testing purposes, the Generic Detector (used for the Seeding and Vertexing optimisation) and the Open Data Detector (ODD)\cite{corentin_allaire_2022_6445359}. The Generic Detector corresponds to the detector design used in the TrackML challenge\cite{Amrouche2023}; it is a typical all-silicon LHC tracking detector with ten layers of cylinders and disks and has been used as a reference for many developments in track reconstruction. The ODD is an evolved version of the Generic Detector implemented using DD4Hep\cite{Frank_2014}, which also provides all the support structure and cabling of a real detector. This allows us to more accurately study the effect of particle-matter interaction in our proton-proton collision. 

	To study the performance of our tuning, we have used simulated $t\bar{t}$ events (which are the standard for tracking performance evaluation) under conditions similar to the HL-LHC: an energy in the centre of mass of $\sqrt{s}=14$ TeV and 200 additional pile-up vertices per event. The events were generated using Pythia8\cite{10.21468/SciPostPhysCodeb.8}, and only particles with a transverse momentum of more than 1~GeV were considered.  

\section{Optimisation Framework}

	We have looked at different derivative-free approaches for our parameter tuning. At its simplest, those algorithms are given a scoring function and will look for the parameters configurations which optimise it. In the case of tracking, the score is performance-based and quite expensive to compute (efficiency, fake rate...) and will tend to be stochastic. This means that classical function minimisation algorithms such as Minuit\cite{James:2296388} won't be effective. Finally, since it would be extremely complex to differentiate our tracking algorithms, we also need a derivative-free approach for our optimisation.

	In this study, we have looked at two different optimisation frameworks: Or\'ion\cite{xavier_bouthillier_2022_0_2_6}, an asynchronous framework for black-box function optimisation and Optuna\cite{optuna_2019}, an open source software for automatic hyperparameter search. Both can easily be used with some simple Python code which can interface with the ACTS' Python-based jobs that we developed to expose the ACTS algorithms' configuration parameters and outputs. This will allow us to tune the parameters easily and extend this optimisation to other algorithms. Each of those frameworks implements different optimisation algorithms that can result in different speed and physics performances.

	In our test with Or\'ion, we used a random search algorithm. As its name implies, it consists of a random sampling of the parameters space. For a large enough number of trials, configurations close to the optimum should have been visited. This method is slow but should converge, given enough trials. It is currently used to verify the validity of the scoring used in the mapping and the feasibility of its optimisation.
	
	In the case of Optuna (in particular for the seeding and vertexing), we tested a Tree-structured Parzen Estimator (TPE)\cite{NIPS2011_86e8f7ab} algorithm. The TPE is a Bayesian optimisation method that builds a probabilistic model and uses it to decide which parameter value to use in the next iteration.
	
\section{Material Mapping}

	When reconstructing the trajectory of a particle, one needs to account for the material in the detector properly. As particles interact with matter, they deviate from their original trajectory; we thus need to increase the search window for hits in the next layer while reconstructing tracks. For this effect to be accounted for, the tracking framework needs to know how much material is present at each point in the detector. Usually, very precise simulations of our detectors exist (most of the time based on Geant4\cite{AGOSTINELLI2003250}), but they are too memory hungry and using them as part of the track reconstruction would greatly slow down the process. 
	
\begin{figure}[h]
\includegraphics[width=17pc]{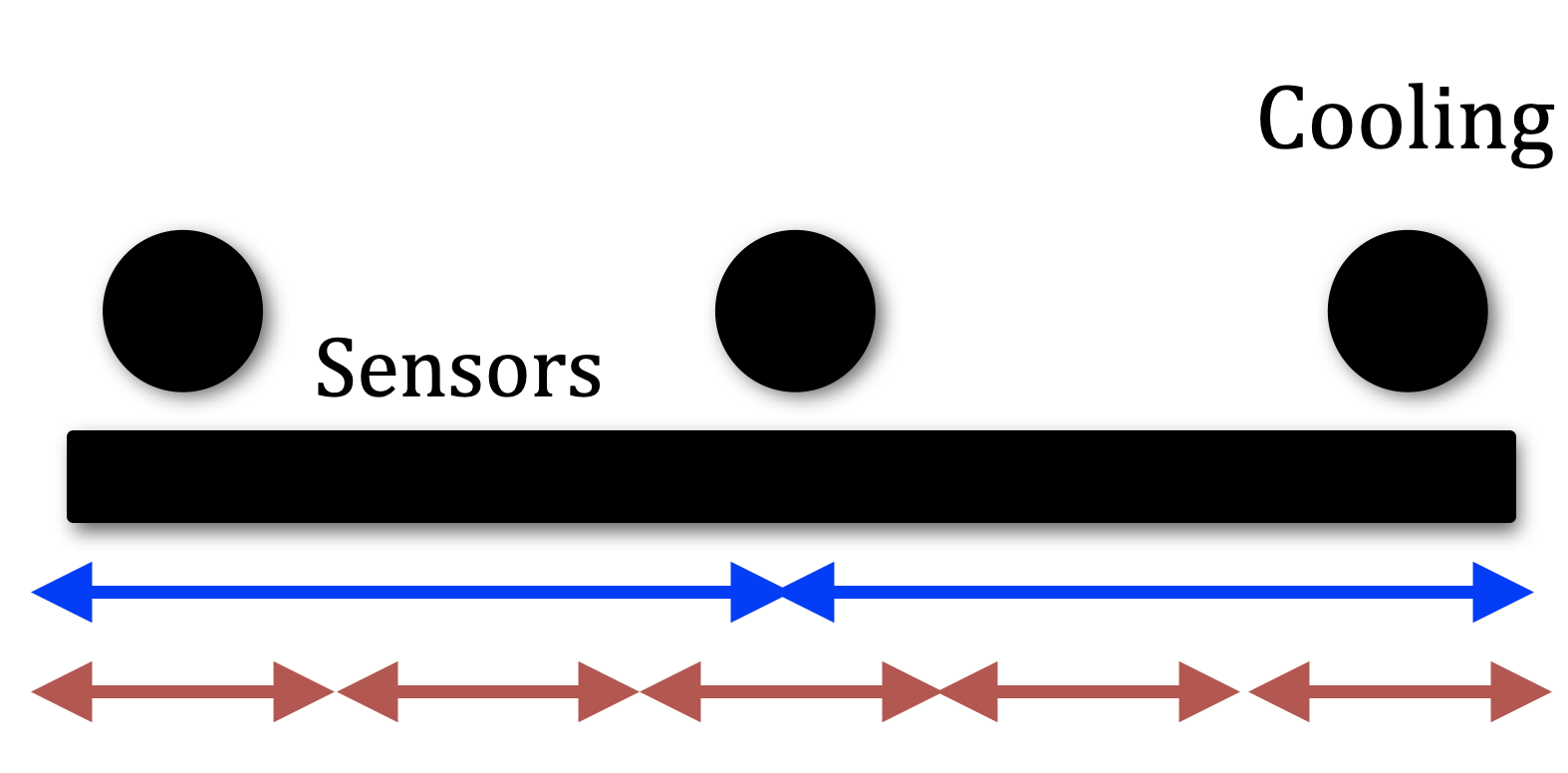}\hspace{4pc}%
\begin{minipage}[b]{16pc}\includegraphics[width=16pc]{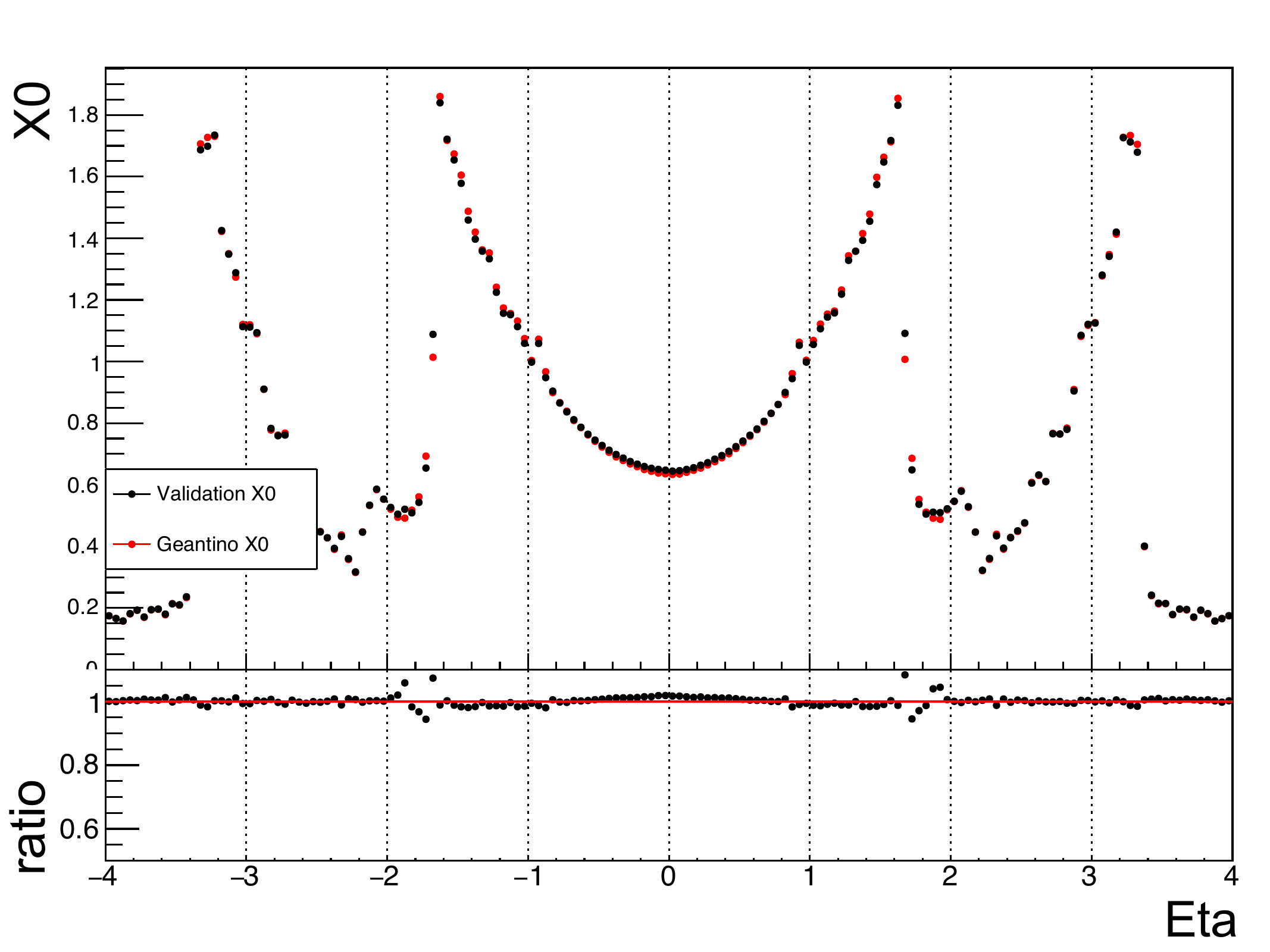}
\end{minipage}
\caption{(a) Projection of the material onto bins (arrows) of different sizes. (b) Comparison between the material encountered in a Geant4 simulation of the ODD and the one in the ACTS propagation.}
\label{Mapping}
\end{figure}

	To solve those issues, we use a simplified material model called a material map that can be used. Those maps are created by projecting all the material in the detector onto a set of predetermined surfaces, usually the entrance surfaces of the different sensitive layers. Each mapping surface is binned in two dimensions, and the material projected in each bin is averaged. When a track intersects one of the mapping surfaces, we determine which bin it crosses, and we compute the material interaction effect based on the material stored in the bin. An illustration of the mapping for one surface can be seen in Figure~\ref{Mapping}(a)
	
	While this method is very effective, producing the map can be quite a long process and require a lot of manual optimisation. We must select which surface we want to map the material onto (usually self-evident) and which binning to apply to each surface. That second part requires a good understanding of the detector geometry and much trial and error. If we choose a binning with too coarse bins, we will miss some of the geometry details, resulting in an especially biased reconstruction. While if we select a binning that is too fine, the time needed to generate the map will increase significantly, and the size in memory of the map will be too large, slowing down its readout.

	Using Or\'ion, we have tried to perform the mapping of the ODD automatically. The user only selects a set of surfaces to map the material onto, and the algorithm finds the optimal binning automatically. For our test, 107 surfaces were used in the mapping, resulting in a total of 214 parameters to optimise. This was performed using a random search algorithm; the focus is finding a score that can be used to judge the quality of a map. After many trials, we decided to use equation~\ref{Score_mat} for the score, where $bins$ is the number of bins in a surface, and $variance$ is the variance of all the material projected onto a given bin. By minimising this score, we keep the number of bins as small as possible while minimising the variance (and thus having bins that adequately represent the local material). This method has been tested and performs quite well; results were obtained after one day on 40 CPU cores and show a good agreement between the material in the map and the simulation Figure~\ref{Mapping}(b).

\begin{equation}
Score = \frac{1}{bins} \times \sum_{bin}variance_{bin} \times (1+\sqrt{bins})
\label{Score_mat}	
\end{equation}

\section{Track Seeding}

	Track finding is a complex problem due to the very large combinatorics that arises when many hits are present in the detector. To simplify this problem, we perform a first step called seed finding, in which we explore hits combinations in the first few layers of the detector. With this exploration, we generate seeds, possible track candidates, that will be used as input for the track finding. In ACTS, seeds consist of triplets of hits in the detector, to which we can apply a helicoidal fit to get a coarse estimation of the corresponding track's parameters. Once this has been done, seeds are filtered using user-defined parameters before being passed to the track finding. Those parameters can differ significantly depending on the detector geometry and experimental conditions; we can thus try to apply our auto-tuning solution to them.
	
\begin{figure}[h]
\includegraphics[width=16pc]{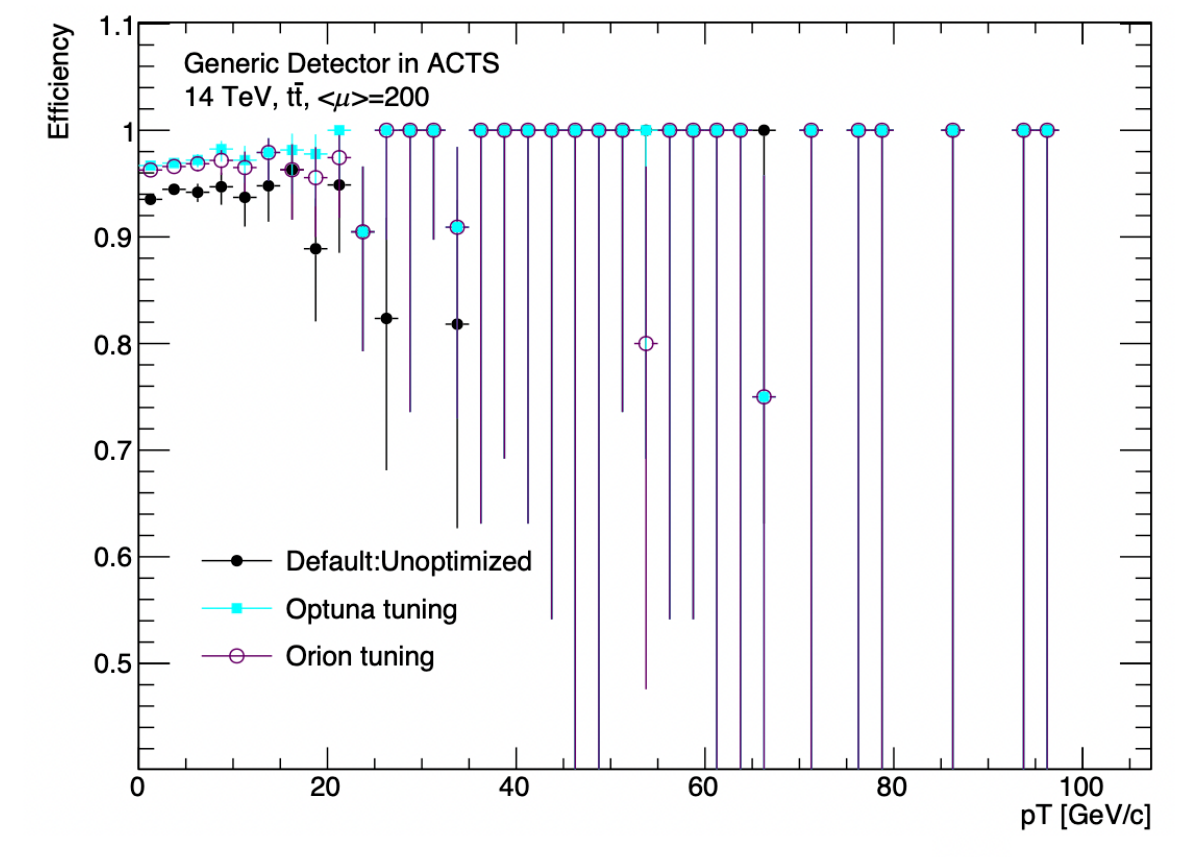}\hspace{4pc}%
\begin{minipage}[b]{17.5pc}\includegraphics[width=17.5pc]{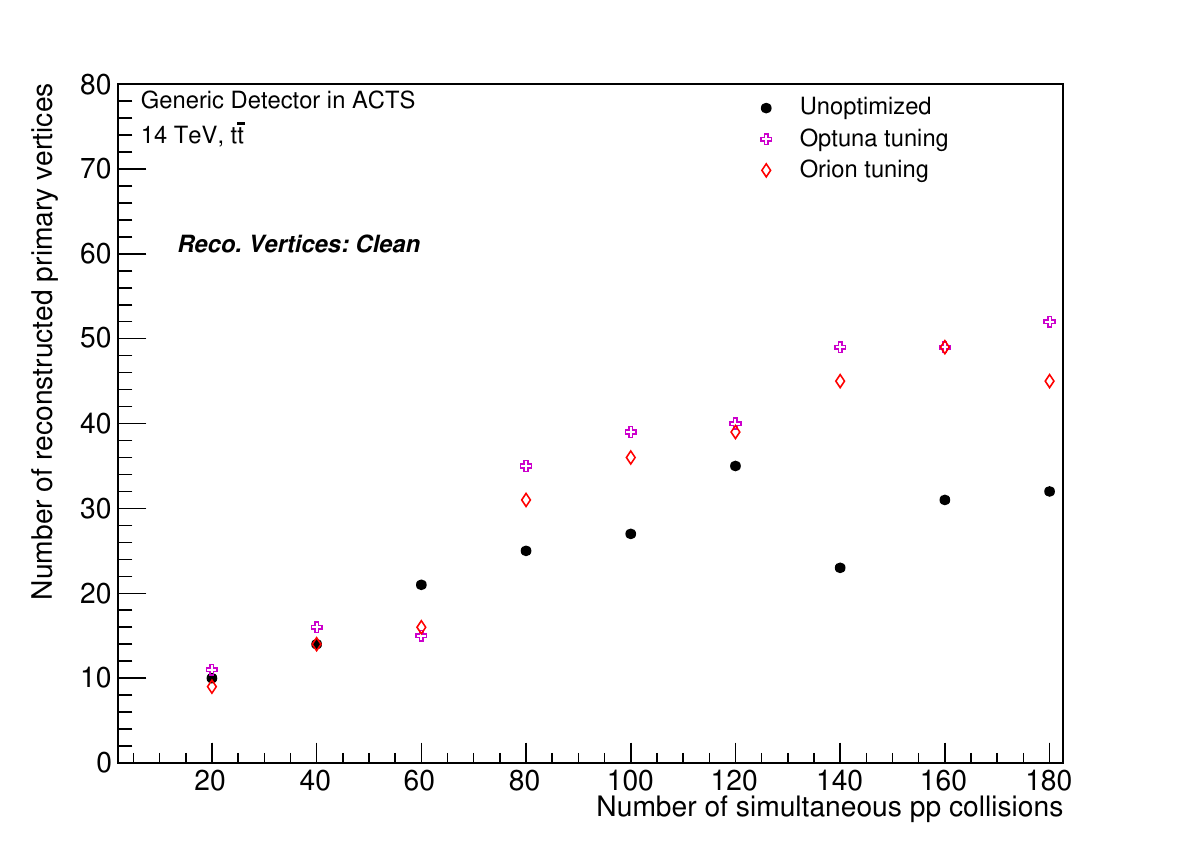}
\end{minipage}
\caption{(a) Track reconstruction efficiency after seeding optimisation. (b) The number of clean vertices reconstructed after optimisation of the vertexing.}
\label{Seeding_Vertexing}
\end{figure}

	It is important to understand that if for a given particle no seed is reconstructed, then the particle is lost. Seeding thus has a significant impact on the reconstruction performance. To study the performances of seeding, we will look at three figures of merit at the end of the tracking chain: the efficiency (fraction of particles reconstructed), the fake rate (fraction of tracks not corresponding to any particle) and the duplicate rate (fraction of tracks that are duplicates of already reconstructed tracks). The first two directly impact the physics performance of the algorithm, while the third will impact the speed of the reconstruction. When tuning the seeding parameter, we will thus use the score from equation~\ref{Score_seed}.

\begin{equation}
Score = Efficiency - (FakeRate + \frac{DuplicateRate}{K} + \frac{RunTime}{K})
\label{Score_seed}	
\end{equation}
	To tune the seeding, we have determined eight relevant parameters that can be optimised. Both Optuna and Or\'ion converge in one hour to a good parameter configuration. The resulting efficiency after track reconstruction is shown in Fig~\ref{Seeding_Vertexing}(a), demonstrating an improvement concerning an unoptimised configuration in both cases. Similar improvements were observed concerning the duplicate and fake rates. 

\section{Vertexing}

	Once tracks have been reconstructed, it is helpful for high pile-up experiments to reconstruct the primary vertices particles originated from. This is performed by the vertexing algorithm. In ACTS, we use an adaptive multi-vertex finder (AMVF)\cite{ATL-PHYS-PUB-2019-015}. This algorithm simultaneously fits all the tracks in the detector, assigning them to the different vertex seeds until all vertices have been fitted. 

	Many variables are used in the evaluation of the performances of this AMVF. The efficiency (fraction of truth vertices reconstructed) and the fake rate (fraction of reconstructed vertices not associated with a truth vertex) describes the capacity of our algorithm to reconstruct vertices. We can then separate the reconstructed vertices into three categories: the clean vertex (associated with only one truth vertex), the merged vertex (associated with multiple truth vertices) and the split vertices (multiple vertices associated with the same truth vertex). Our goal is to reconstruct as many clean vertices as possible while minimising poorly reconstructed ones. Using the efficiencies and the fraction of vertices in each category, we can compute a score representing the quality of the vertex reconstruction; this score is shown in equation~\ref{Score_vertex}.

\begin{equation}
Score = (Eff_{Total}+2Eff_{Cleaned})-(Merged+Split+Fake+Resolution)
\label{Score_vertex}	
\end{equation}

	The optimisation algorithm converged in roughly four hours to optimise the five parameters of the vertexing. The resulting number of clean vertices is shown as a function of the pile-up in Figure~\ref{Seeding_Vertexing}(b). Good improvement in the number of cleaned vertices can be seen concerning the unoptimised configuration, especially at high pile-up. Similarly, a reduction in the number of fakes at high pile-ups was observed.

\section{Conclusion}

	We have shown that data-driven auto-tuning algorithms can be used in the context of track reconstruction to optimise the input parameters of different algorithms such as seed reconstruction, vertex reconstruction and material mapping. Those methods have been implemented in the ACTS framework and can thus be used by any experiment using it for their tracking needs. In the future, our effort will be directed toward generalising this approach so that most algorithms in ACTS can be automatically tuned. 

\section{Acknowledgments}
This project has received funding from the European Union’s Horizon 2020 research and innovation programme under grant agreement No 101004761. \newline
This work was supported by the National Science Foundation under Cooperative Agreement OAC-1836650.

\section*{References}
\bibliography{Bibliographie}

\end{document}